\documentclass{aa}
\usepackage{amssymb}
\usepackage{amsmath}
\usepackage{xspace}
\usepackage{graphicx}
\usepackage{enumerate}   
\usepackage{color}
\usepackage{hyperref}

\usepackage{natbib}

\def\gr{$\gamma$-ray}
\def\fermi{Fermi\xspace}

\usepackage{color}

\begin{document}

\title{Cosmic-ray spectrum in the local Galaxy}
\author{ Andrii Neronov$^1$, Denys Malyshev$^2$, Dmitri V. Semikoz$^{3,4}$} 
\institute{$^1$Astronomy Department, University of Geneva, Ch. d'Ecogia 16, 1290, Versoix, Switzerland\\
$^2$ Institut f{\"u}r Astronomie und Astrophysik T{\"u}bingen, Universit{\"a}t T{\"u}bingen, Sand 1, D-72076 T{\"u}bingen, Germany\\
$^3$ APC, Universite Paris Diderot, CNRS/IN2P3, CEA/IRFU,
Observatoire de Paris, Sorbonne Paris Cite, Paris, France\\
$^4$National Research Nuclear University MEPHI (Moscow Engineering Physics Institute),  Moscow, Russia
}

\authorrunning{Neronov et al.}
\titlerunning{Nature of the low-energy break}
\abstract
{}
{
We study the spectral properties of the cosmic-ray spectrum in the interstellar medium within 1~kpc distance from the Sun.   
 }
{ 
We used eight-year exposure data of molecular clouds of the Gould Belt obtained with the Fermi/LAT telescope to precisely measure the cosmic-ray spectrum at different locations in the local Galaxy.  We compared this measurement with the direct measurements of the cosmic-ray flux in and around the solar system obtained by Voyager and AMS-02 or PAMELA.  
 }
{
We find that the average cosmic-ray spectrum in the local Galaxy  in the 1-100 GeV range is well described by a broken power-law in rigidity with a low-energy slope of $2.33^{+0.06}_{-0.08}$ and a break at $18^{+7}_{-4}$ GV, with a slope change by $0.59\pm 0.11$. This result is consistent with an earlier analysis of the \gr\ signal from the Gould Belt clouds based on a shorter exposure of Fermi/LAT  and with a different event selection.  The break at 10-20 GV is also consistent with the combined Voyager + AMS-02 measurements in/around the solar system. The slope of the spectrum below the break agrees  with the slope of the average cosmic-ray spectrum in the inner part of the disk of the Milky Way that was previously derived from the Fermi/LAT \gr\ data.  
%We put forward a conjecture that it is this slope (rather than conventionally assumed softer slope 2.7 ... 2.8) that is determined by the balance of steady state injection of cosmic rays with the powerlaw spectrum with the slope 2 ... 2.1 by Fermi acceleration process and energy-dependent diffusive escape of cosmic ray particles through the turbulent interstellar magnetic field with Kolmogorov turbulence spectrum.
We conjecture that it is this slope of 2.33 and not the locally measured softer slope of 2.7--2.8 that is determined by the balance between a steady-state injection of cosmic rays with a power-law slope of 2--2.1 that is due to Fermi acceleration and the energy-dependent propagation of cosmic-ray particles through the turbulent interstellar magnetic field with a Kolmogorov turbulence spectrum.
The approximation of a continuous-in-time injection of cosmic rays at a constant rate breaks down, which
causes the softening of the spectrum at higher energies. 
}
{}


\maketitle
%%%%%%%%%%%%%%%%%%%%%%%%%%%%%%%%%%%%%%%%%%%%%%%%%
\section{Introduction}
%%%%%%%%%%%%%%%%%%%%%%%%%%%%%%%%%%%%%%%%%%%%%%%%%

Our understanding of the injection and propagation of Galactic cosmic rays is to a great part based on the measurement at a single point in the Milky Way: the location of the solar system. Even this measurement has been strongly influenced by the effect of solar modulation, which distorts the low-energy part of the spectrum around GeV. This has changed with the new Voyager measurement, which was obtained in the interstellar medium outside the boundary of the solar system \citep{stone,cummings}. A comparison of the Voyager measurement with measurements made at higher energies where the signal is not affected by the solar modulation \citep{ams-02_proton,pamela_proton} show that the spectrum in the 1-10~GV range does not follow the approximate power law in rigidity shape that is found at high energies. A modelling of the spectrum by \cite{cummings} introduces a range of breaks in the spectrum of cosmic-ray injection and in the energy dependence of the cosmic-ray diffusion coefficient to explain the change in behaviour of the spectrum.   In a similar way, modelling of the Voyager and AMS-02 data based on the HelMod  solar modulation and GALPROP cosmic-ray propagation numerical models finds that the injection spectrum of cosmic rays has to have a break in the 10 GV range \citep{boschini17} as well as a number of features introduced by the re-acceleration in the interstellar medium and changes in the energy dependences of the cosmic-ray diffusion coefficient. In contrast, the analysis of \citet{voyager_pamela_ams} found that only one break at $\sim 10$~GV is found to be sufficient in the phenomenological broken power-law description of the interstellar cosmic-ray spectrum in 1-100 GeV range.  With such uncertainties in the measurements, an unambiguous physical interpretation of the observed spectral slope change(s) in the 10 GeV range appears problematic.  

%%%%%%%%%%%%%%%%%%%%%%%%%%%%%%%%%%%%%%%%%%%%%%%%%
\begin{table*}
\begin{center}
\begin{tabular}{|c|c|c|c|c|c|}
\hline
Name  & $(l_s,b_s)$ & $\theta$,deg & $(l_b,b_b)$ & $D$, pc & $M/10^5M_\odot$  \\
\hline
R CrA   & (0.56 ; -19.63) & 6 & (6.94 ; -19.63)  &  150 & 0.03  \\
%& & & (353.69 ; -19.63) &   & &&& \\
Rho Oph  & (355.81;16.63) & 10 & (34.94;16.63) &  165 & 0.3 \\
Perseus   & (159.31; -20.25) & 8 & (148.44; -19.88) &  350 & 1.3 \\
Chameleon   & (300.43 ; -16.13) & 11 & (283.81; -16.13) &  215 & 0.1 \\
Cepheus   &(108.56; 14.75) & 12 &(85.94; 14.75) & 450  & 1.9  \\
Taurus   & (173.19; -14.75) & 12  & (143.94; -19.50) & 140  & 0.3 \\
Orion A   & (212.19; -19.13) & 8  & (233.69; -19.13) &  500  & 1.6 \\
Orion B   & (204.56; -13.75) & 7.25 & (233.69; -19.13) & 500 & 1.7   \\
Mon R2   & (213.81; -12.63) & 3 & (233.81; -18.75) & 830  & 1.2  \\
\hline
\end{tabular}
\end{center}
\caption{Parameters of the clouds used for the analysis, including galactic coordinates of the centres of the box-shaped templates for the cloud $(l_s,b_s)$ and background-estimation region $(l_b;b_b)$, the side of the box $\theta$. Distances $D$ and masses $M$ are from \citet{co}. }
\label{tab:clouds_properties}
\end{table*}
%%%%%%%%%%%%%%%%%%%%%%%%%%%%%%%%%%%%%%%%%%%%%%%%%

The cosmic-ray spectrum at locations different from the solar system was measured based on \gr\ observations of nearby molecular clouds. Such clouds form  the Gould Belt, which is a ring-like structure around the solar system with a diameter of about 1~kpc \citep{gouldbelt}. The \gr\ measurements indicate a break in the cosmic-ray spectrum in the 10~GeV energy range \citep{lowenergybreak}. It is possible that the break  or turnover in the cosmic ray spectrum near the solar system corresponds to the break inferred from the \gr\ measurements of the Gould Belt clouds. The measurement
precision of the cosmic ray spectrum with this method is affected by the possible presence of an electron Bremsstrahlung component of the \gr\ flux at low energy \citep{lowenergybreak}, possible effects of cosmic-ray propagation inside the molecular clouds \citep{yang}, and uncertainties of the \gr\ yield of pion production/decay reaction \citep{dermer12,kachelriess12}.  

Still another measurement of the cosmic-ray spectrum at more distant locations in the Galaxy is derived from the diffuse \gr\ emission from the Galactic disk. This measurement indicates that the cosmic-ray spectrum in the inner Galactic disk (within the solar orbit) is characterised by a harder slope, $dN/dE\propto  E^{-\Gamma}$ with $\Gamma\simeq 2.4 ... 2.5$ \citep{hard,yang,hard_fermi}, compared to the locally measured spectral slope $\Gamma \simeq 2.8 ... 2.85$ \citep{pamela_proton,ams-02_proton}. The origin of the difference in the slopes of the local and average Galactic disk spectra is not clear. It is possible that the local spectrum is affected by the history of recent star formation activity in the local interstellar medium \citep{lowenergybreak,local_source,savchenko15}. Otherwise, the change in slope could be due to the changes in the turbulent component of Galactic magnetic field, resulting in the gradually changing energy slope of the cosmic-ray diffusion coefficient \citep{gaggero}.

In this paper we follow up on the analysis of \citet{lowenergybreak} using longer exposure and updated calibrations of the Fermi Large Area Telescope (LAT). This allows us to derive tighter constraints on the shape of the cosmic ray spectrum and to clarify the relation between the locally measured spectrum and the average Galactic disk cosmic-ray spectrum.

%%%%%%%%%%%%%%%%%%%%%%%%%%%%%%%%%%%%%%%%%%%%%%%%%
\section{Data selection and data analysis}
%%%%%%%%%%%%%%%%%%%%%%%%%%%%%%%%%%%%%%%%%%%%%%%%%

Our analysis uses 8.4 years of Fermi/LAT data collected in the period between August 4, 2008 and January 16, 2017 (compare with the three-year exposure considered by \cite{lowenergybreak}). We performed a binned analysis\footnote{see \url{https://fermi.gsfc.nasa.gov/ssc/data/analysis/scitools/extended/extended.html} for a detailed description} of the data in the 0.08 -- 300~GeV energy band for the ``CLEAN'' class photons with the \texttt{P8R2\_CLEAN\_V6} response functions\footnote{see \url{https://fermi.gsfc.nasa.gov/ssc/data/analysis/documentation/Cicerone/Cicerone\_LAT\_IRFs/IRF\_overview.html} }. The analysis is based on fitting the model of the region of interest (ROI) to the data. The molecular clouds considered for the analysis are the same as were analysed by \cite{lowenergybreak}. In order to cover the broad Fermi/LAT point-spread function at $\sim 100$~MeV energies, we selected the ROI radii to be $20^\circ$. Information on the analysed clouds is given in Table~\ref{tab:clouds_properties}.

In each individual case, the ROI model included all sources from the four-year Fermi/LAT 3~FGL catalogue~\citep{fermi_cat} and templates for the molecular cloud and galactic and extra-galactic (\texttt{iso\_P8R2\_CLEAN\_v06.txt}) diffuse emissions. The catalogue sources were assumed to be described in each narrow energy bin by a simple $E^{-2}$ type spectral model and with normalisations that were allowed to vary in every considered energy bin. Within the analysis, the model flux was convolved with the point-spread function and exposure at given sky coordinates and fitted to the observed count map. The best-fit source parameters directly correspond to the expected source flux in the considered energy bin. Where applicable, the upper limits were calculated with
the \texttt{UpperLimits} python module for best-fit test statistic $TS<4,$ and they correspond to a 95\% $\simeq 2\sigma$ probability of the flux to be lower than specified.

All sources considered for the analysis are extended, spanning several degrees on the sky, and in some cases, they are located in crowded regions near the Galactic plane. The \gr\ emission from the Gould Belt clouds is included by default in the diffuse Galactic background model used in the LAT data analysis. To analyse the signal from the clouds, we therefore need to remove the cloud component from the Galactic diffuse emission model template (\texttt{gll\_iem\_v06.fits} was considered in our analysis). The diffuse emission template consists of a set of all-sky maps representing the expected background emission at given energies. To remove the cloud signal, we replaced in each energy band  the template emission in a box around the cloud with a constant determined as a mean over a box of the same size located in nearby background region (see Table~\ref{tab:clouds_properties} for the corresponding parameters). We used the new diffuse emission templates obtained in this way to model the Galactic diffuse emission. In order to produce the template for the molecular cloud emission, we subtracted the refined template for the Galactic diffuse emission from the standard template. We then used the new templates and the spatial models of the clouds in the standard likelihood analysis. Following the recommendations of the Fermi/LAT team, we performed the analysis with enabled energy dispersion handling\footnote{\url{http://fermi.gsfc.nasa.gov/ssc/data/analysis/documentation /Pass8\_edisp\_usage.html}}. 
%%%%%%%%%%%%%%%%%%%%%%%%%%%%%%%%%%%%%%%%%%%%%%%%%
\begin{figure}
\includegraphics[width=\linewidth]{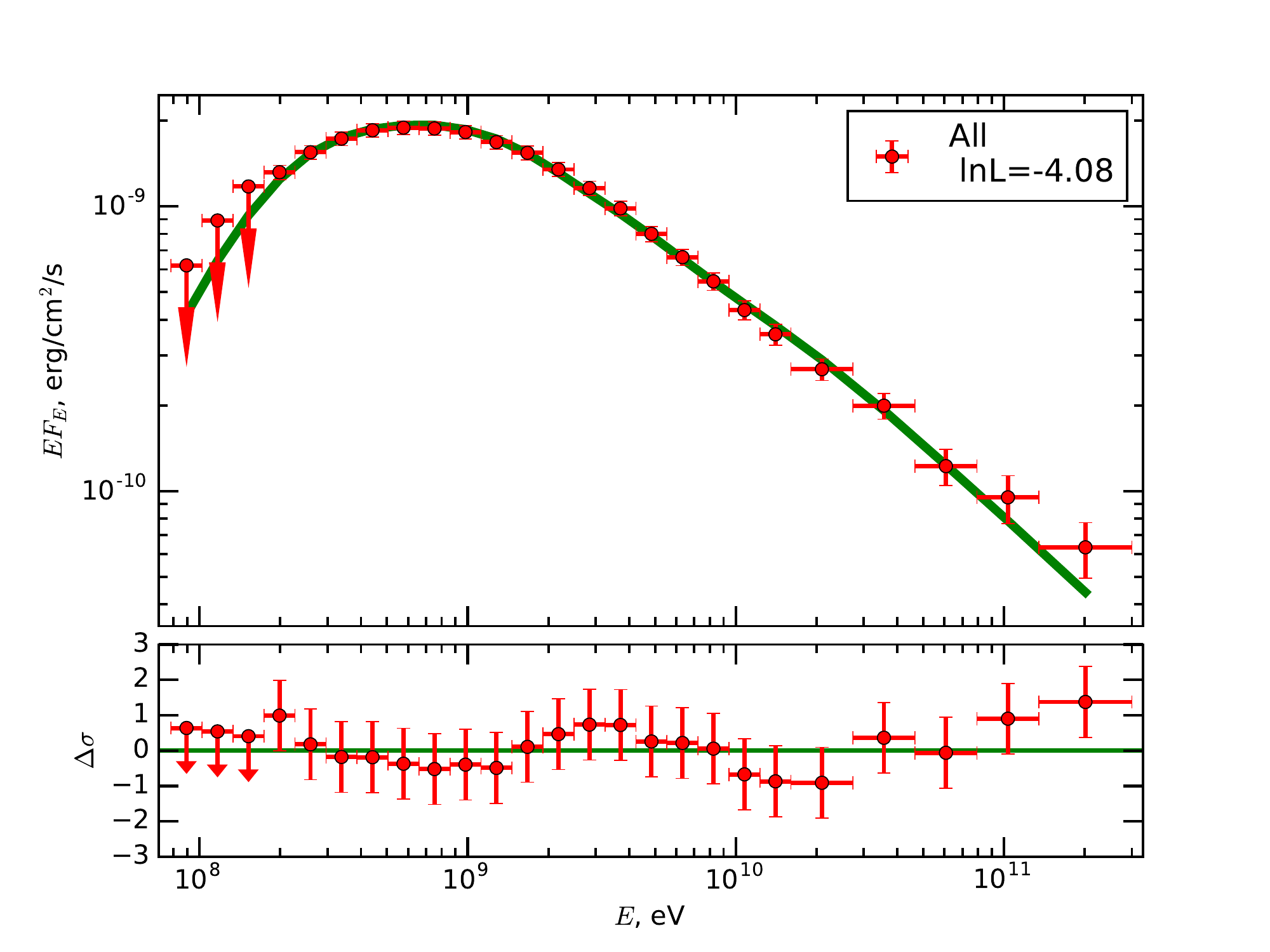}
\caption{Stacked energy spectrum of all the clouds. The solid green line shows the best-fit model for  a broken power-law cosmic-ray distribution, see Eq. (\ref{eq:fit_model}). The lower panel shows the significance of the residuals between the data points and the model in standard deviation units. }
\label{fig:staked_spectra}
\end{figure}
%%%%%%%%%%%%%%%%%%%%%%%%%%%%%%%%%%%%%%%%%%%%%%%%%

%%%%%%%%%%%%%%%%%%%%%%%%%%%%%%%%%%%%%%%%%%%%%%%%%
\section{Results}
%%%%%%%%%%%%%%%%%%%%%%%%%%%%%%%%%%%%%%%%%%%%%%%%%

The stacked spectrum of all the clouds (defined as the energy bin-by-bin sum of the spectra of individual clouds; upper limits are treated as points with half of the upper-limit flux and an
uncertainty equal to the attributed flux)  is shown in Fig.~\ref{fig:staked_spectra}.   The measurements include the systematic errors\footnote{  added as described at    \url{https://fermi.gsfc.nasa.gov/ssc/data/ analysis/scitools/Aeff\_Systematics.html }}. Spectral measurements available in the range below 200~MeV are explicitly replaced by $2\sigma$ upper limits. This was done to reduce the possible contamination of the spectral modelling reported below with  a not-accounted-for electron Bremsstrahlung component that is expected to appear in this energy range \citep{lowenergybreak}.

We fitted the spectrum with a model  of \gr\ emission from neutral pion decays using the\footnote{see \url{http://naima.readthedocs.io/en/latest/}} \texttt{naima 0.8} python package \citep{naima} based on the parameterization of pion production/decay cross sections by \citet{naima1}. The distribution function of incident cosmic rays was assumed to be a broken power-law with a rigidity $r$ with the low- and high-energy indexes $i_1$, $i_2$,  normalisation $N_0$, break rigidity $r_{br}$ (break energy $E_{br}$), and the break strength $s$,
\begin{equation}
F(r) = N_0\cdot (r/(1\mbox{~GV}))^{i_1}/(1+(r/r_{br})^{s})^{((i_1-i_2)/s)}
\label{eq:fit_model}
.\end{equation}
Protons and heavier nuclei were assumed to have identical spectra as a function of rigidity. The nuclear contribution to the \gr\ flux was assumed to be well described by the nearly energy-independent nuclear enhancement factor in the range 1.5 ... 1.8 \citep{mori09}.

\begin{figure}
\includegraphics[width=\linewidth]{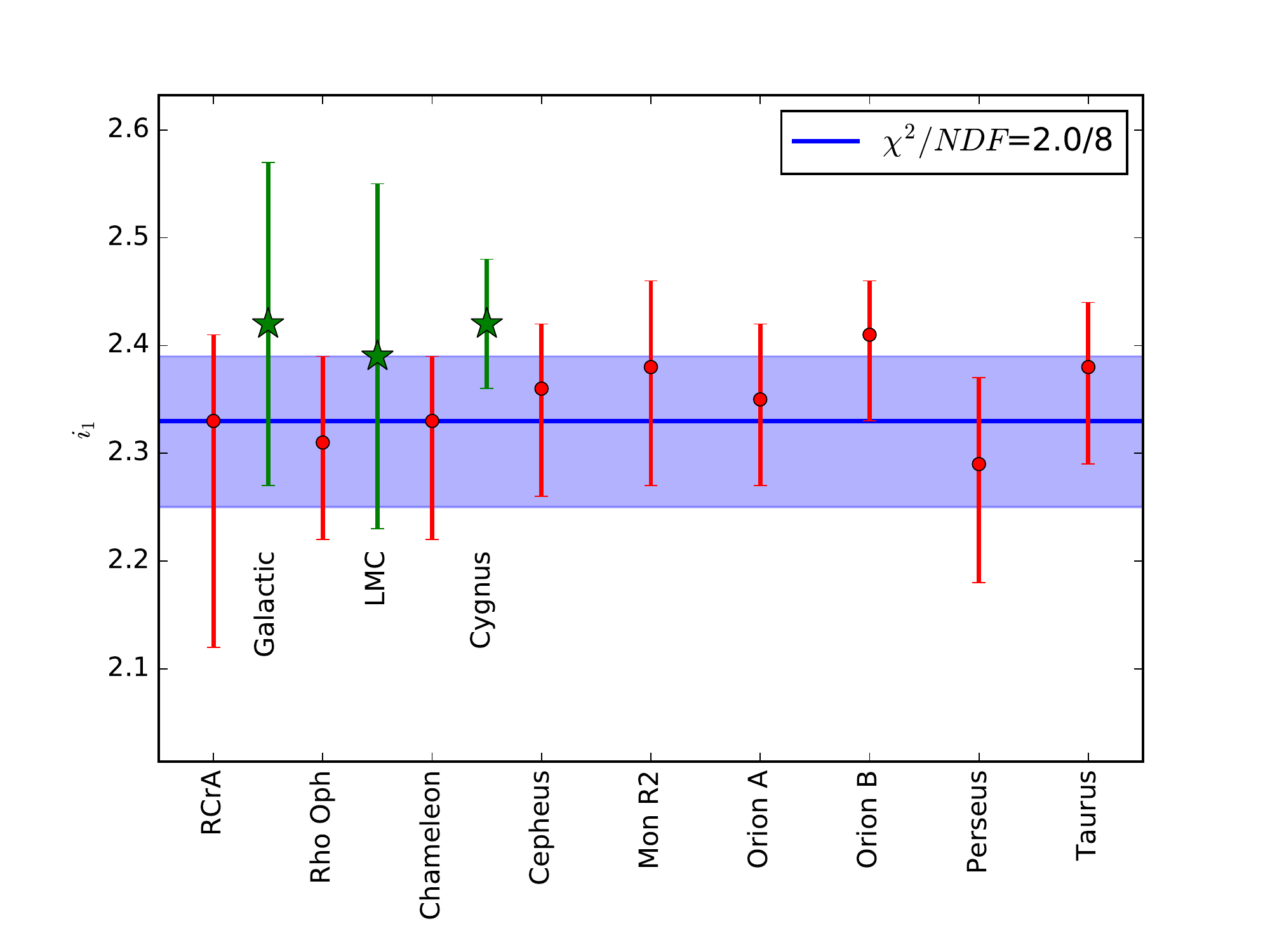}
\caption{ Best-fit low-energy index $i_1$ for individual molecular clouds overplotted with the results from fitting the stacked spectrum. The green point marked with stars shows the slopes observed by~\citet{hard} in the Galactic disk, the LMC, and in
the Cygnus region. }
\label{fig:best_fit_params}
\end{figure}

The model fit parameters for all the clouds are summarised in Table~\ref{tab:fit_params} with the best-fit low energy indexes shown in Fig.~\ref{fig:best_fit_params}. The blue solid line and the shaded region in Fig. \ref{fig:best_fit_params} correspond to the parameters obtained for the fit to the stacked spectrum  of all  clouds (denoted as ``All'' in Table~\ref{tab:fit_params}). The strength of the break is weakly constrained in each individual cloud, and for the stacked spectrum, it can be constrained as $s>1.5$ at 95\% c.l.

%%%%%%%%%%%%%%%%%%%%%%%%%%%%%%%%%%%%%%%%%%%%%%%%%
%%%%%%%%%%%%%%%%%%%%%%%%%%%%%%%%%%%%%%%%%%%%%%%%%

\begin{figure*}
\includegraphics[width=0.32\linewidth]{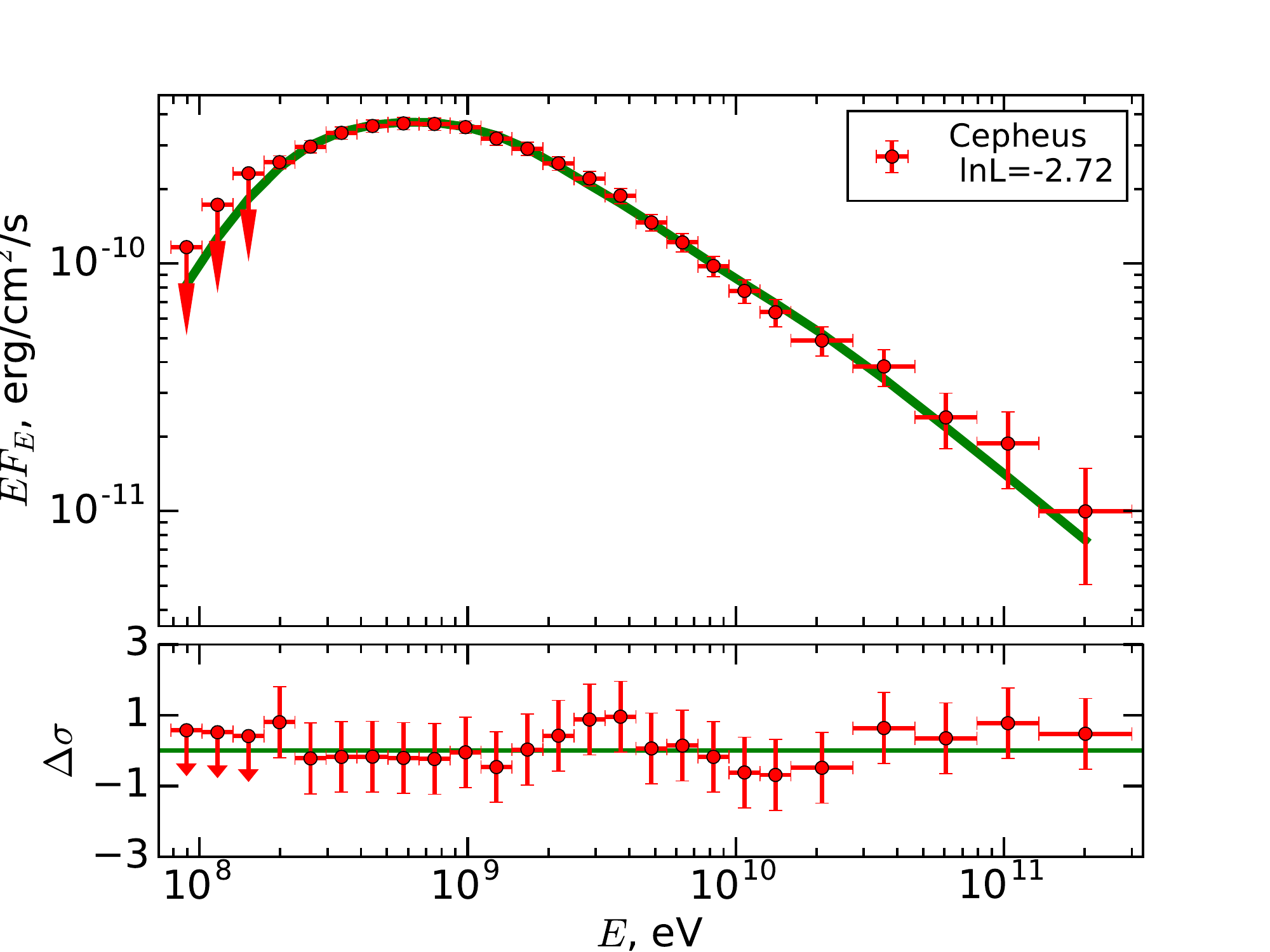}
\includegraphics[width=0.32\linewidth]{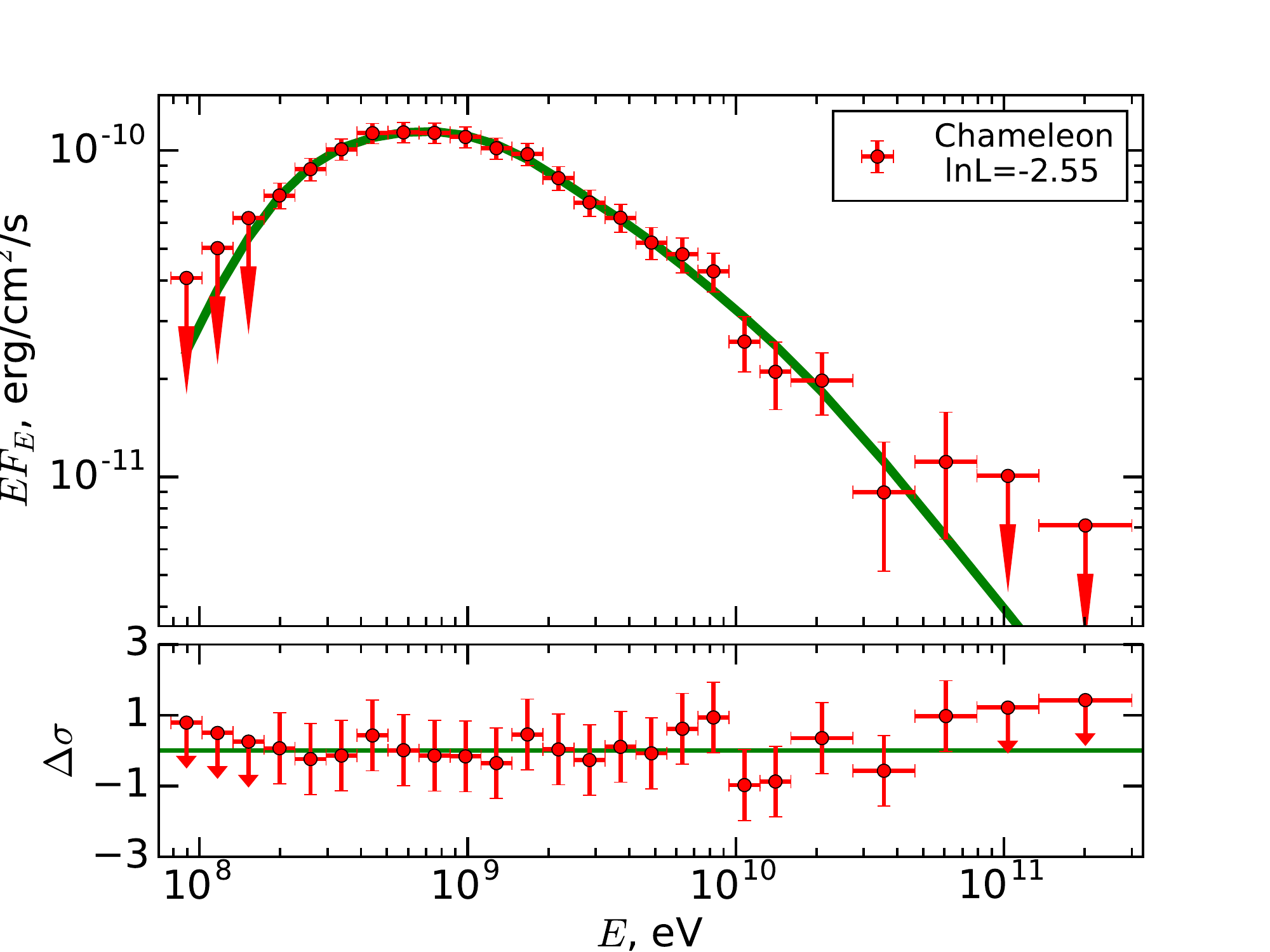}
\includegraphics[width=0.32\linewidth]{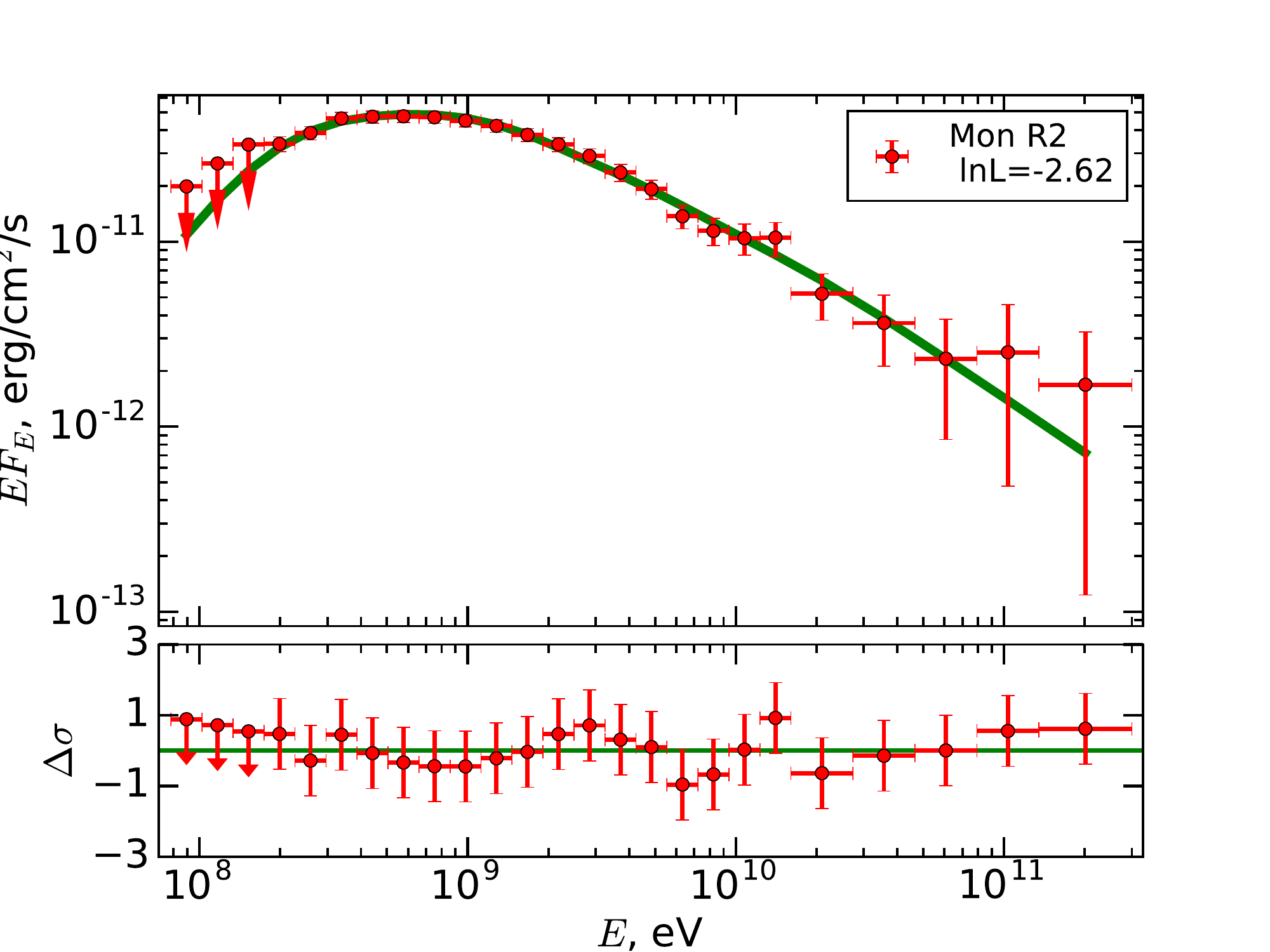}\\
\includegraphics[width=0.32\linewidth]{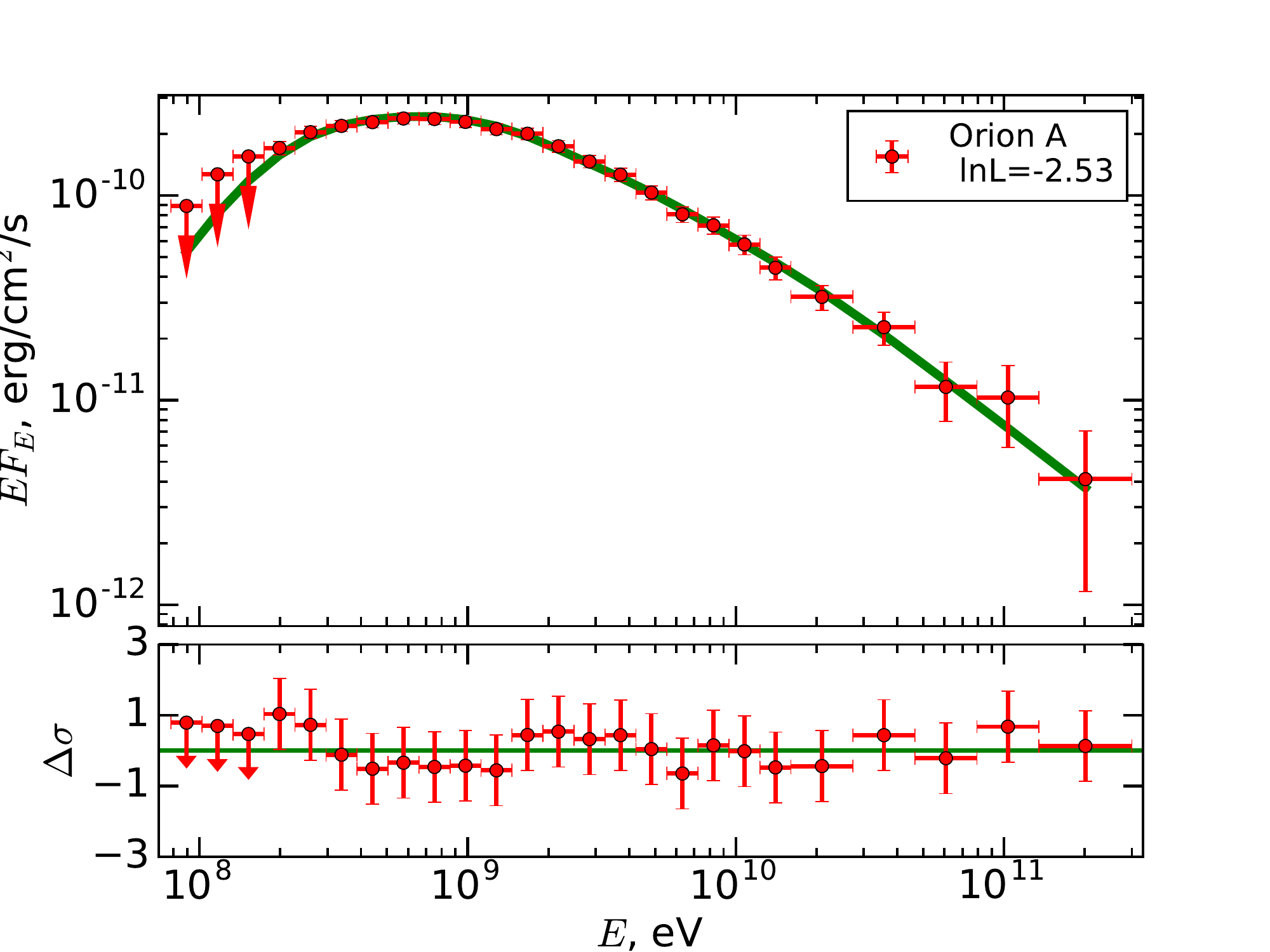}
\includegraphics[width=0.32\linewidth]{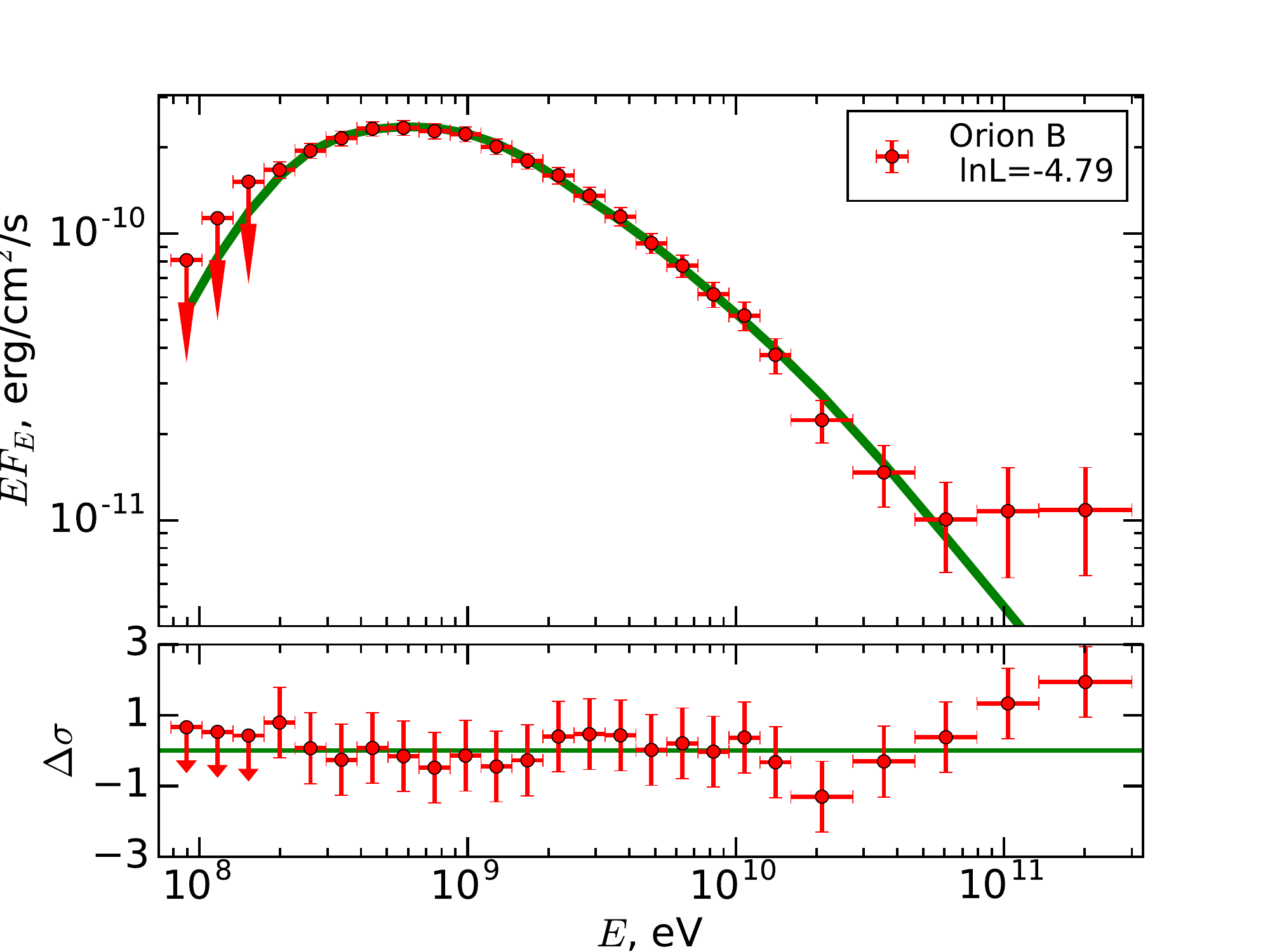}
\includegraphics[width=0.32\linewidth]{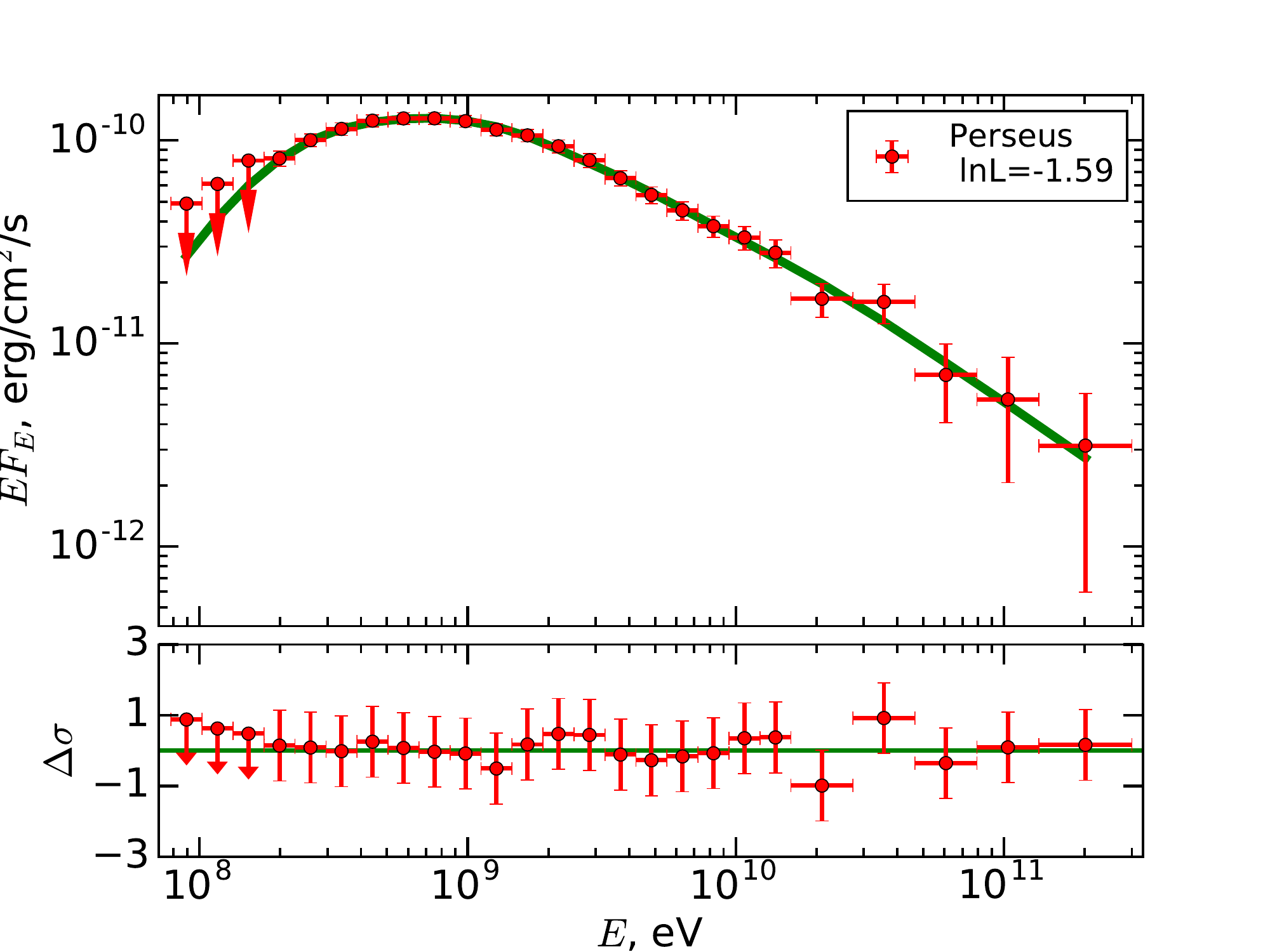}\\
\includegraphics[width=0.32\linewidth]{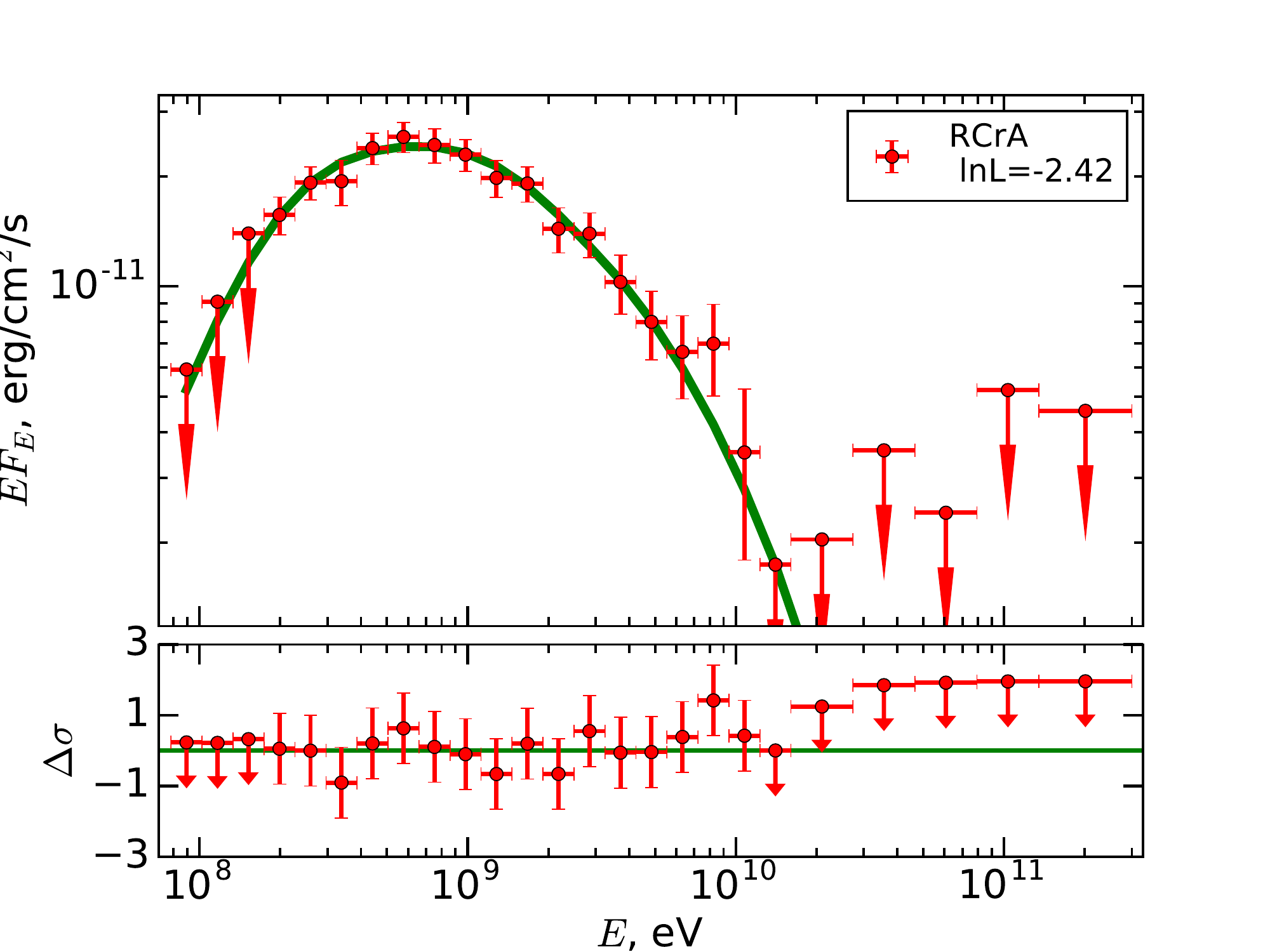}
\includegraphics[width=0.32\linewidth]{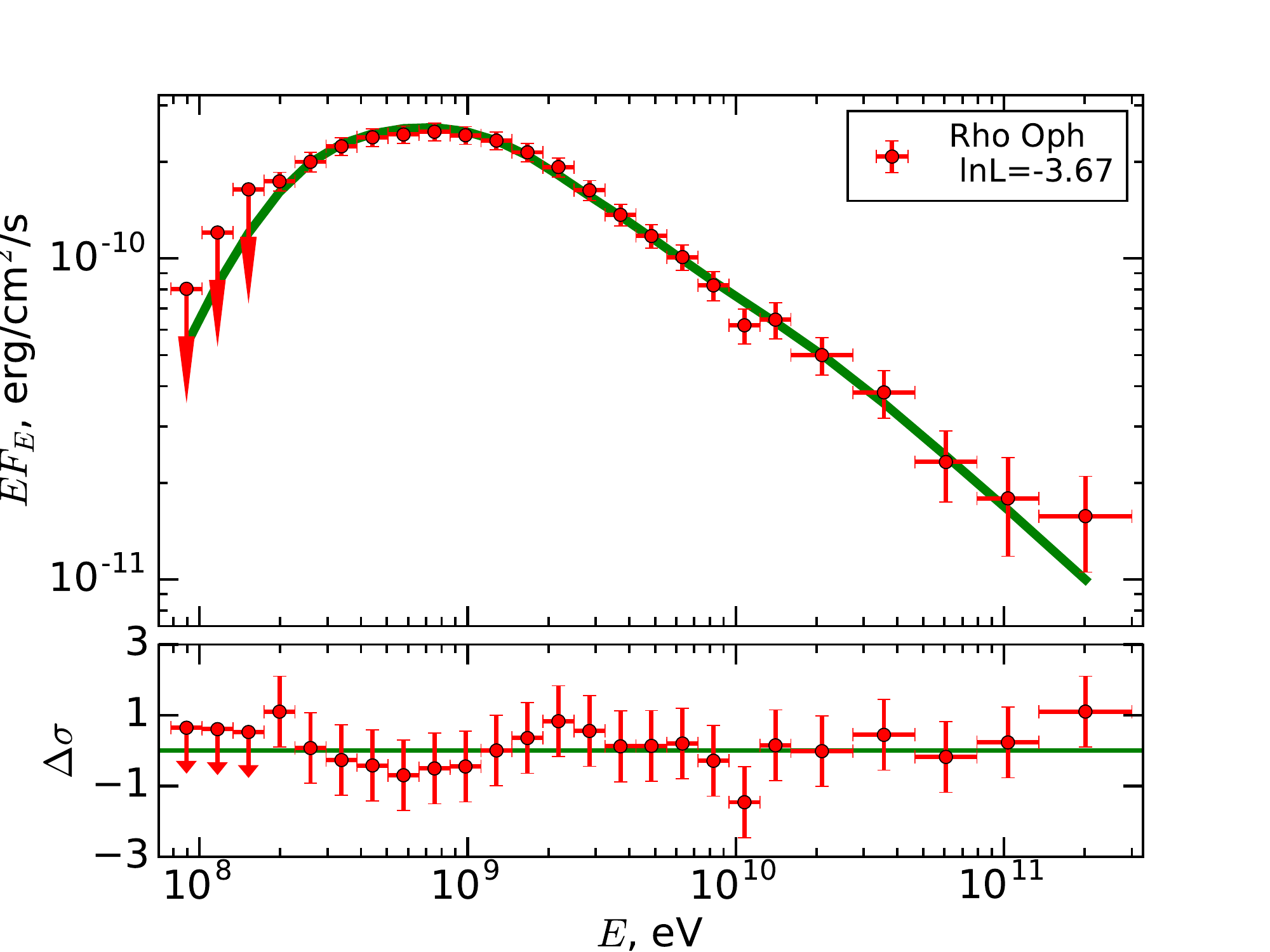}
\includegraphics[width=0.32\linewidth]{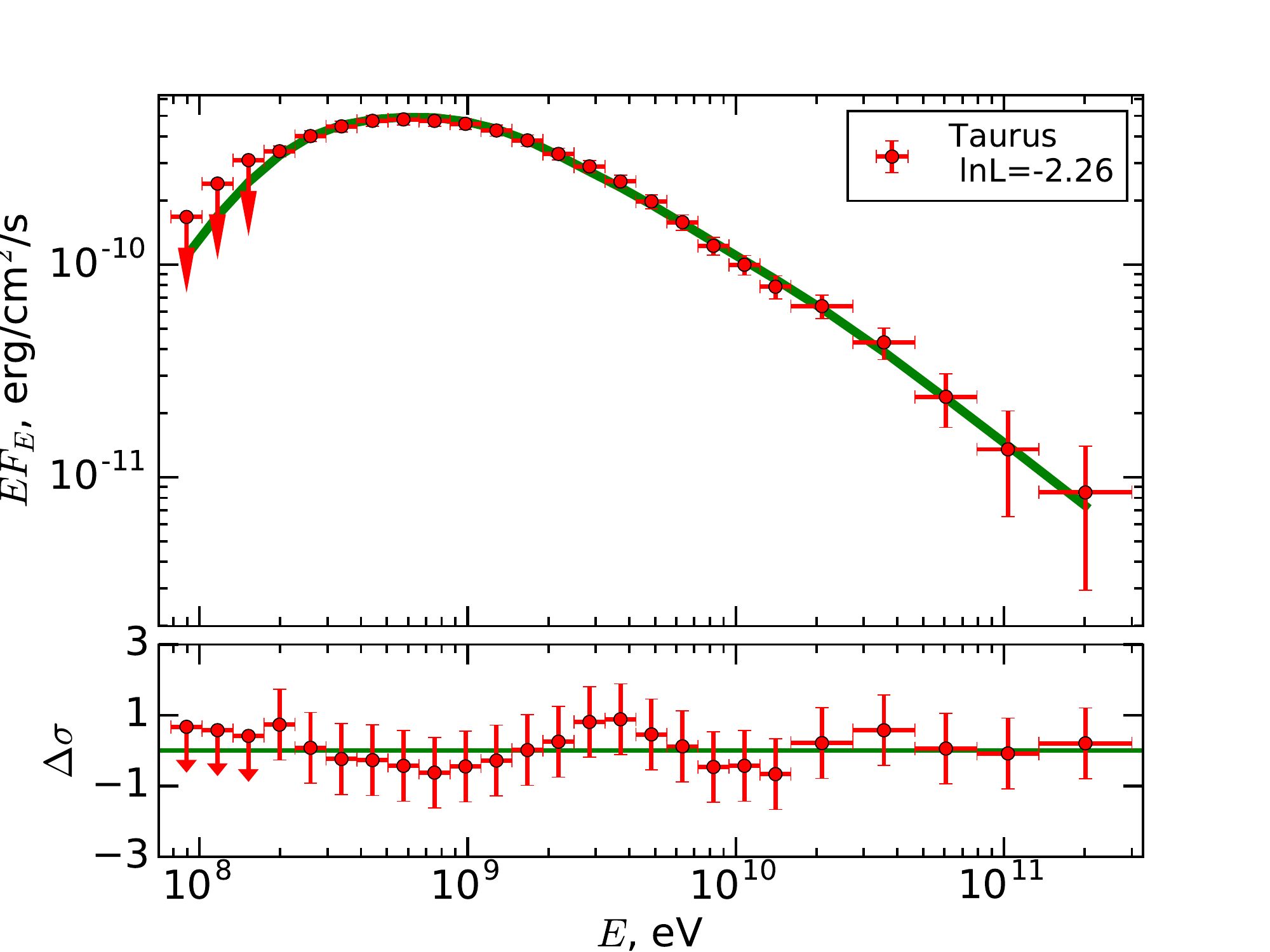}
\caption{\gr\ spectra of molecular clouds from the Gould Belt together with the best-fit models.  The lower panel shows the significance of the residuals between the data points and the model in standard deviation units. }
\label{fig:all_spectra}
\end{figure*}

The spectra of individual clouds are shown in Fig.~\ref{fig:all_spectra}, together with the fit residuals. The broken power-law models clearly provide satisfactory fits to the spectra of all the clouds and also to the stacked cloud spectrum.

%%%%%%%%%%%%%%%%%%%%%%%%%%%%%%%%%%%%%%%%%%%%%%%%%
%%%%%%%%%%%%%%%%%%%%%%%%%%%%%%%%%%%%%%%%%%%%%%%%%

%%%%%%%%%%%%%%%%%%%%%%%%%%%%%%%%%%%%%%%%%%%%%%%%%
\section{Discussion}
%%%%%%%%%%%%%%%%%%%%%%%%%%%%%%%%%%%%%%%%%%%%%%%%%

The updated measurement of the cosmic-ray spectrum in the local Galaxy based on the \gr\ observations of the Gould Belt clouds is broadly consistent with the previous observation of the break that was based on a shorter Fermi/LAT exposure \citep{lowenergybreak}, after re-modelling of the spectrum as a broken power-law in rigidity rather than broken power-law in kinetic energy. The results of \citep{lowenergybreak} imply a low-energy slope $i_1\simeq 2.4$ of the broken power-law in rigidity, as previously noted by \citet{nath12}, and the high-energy indices are also consistent. The difference between the measurement of the break energy/rigidity ($E_{br}=9\pm 3$ GeV in \citet{lowenergybreak}) is explained by (a)  the systematic error, which is related to our different knowledge of the  calibrations of  Fermi/LAT in the event selections\footnote{Pass 7 data selection used in the analysis of \citet{lowenergybreak} vs. Pass 8 data selection of the present analysis.} , and (b) the difference in the shapes of the two models (energy or rigidity power laws) in the GeV range. 

\begin{figure*}
\includegraphics[width=0.5\linewidth]{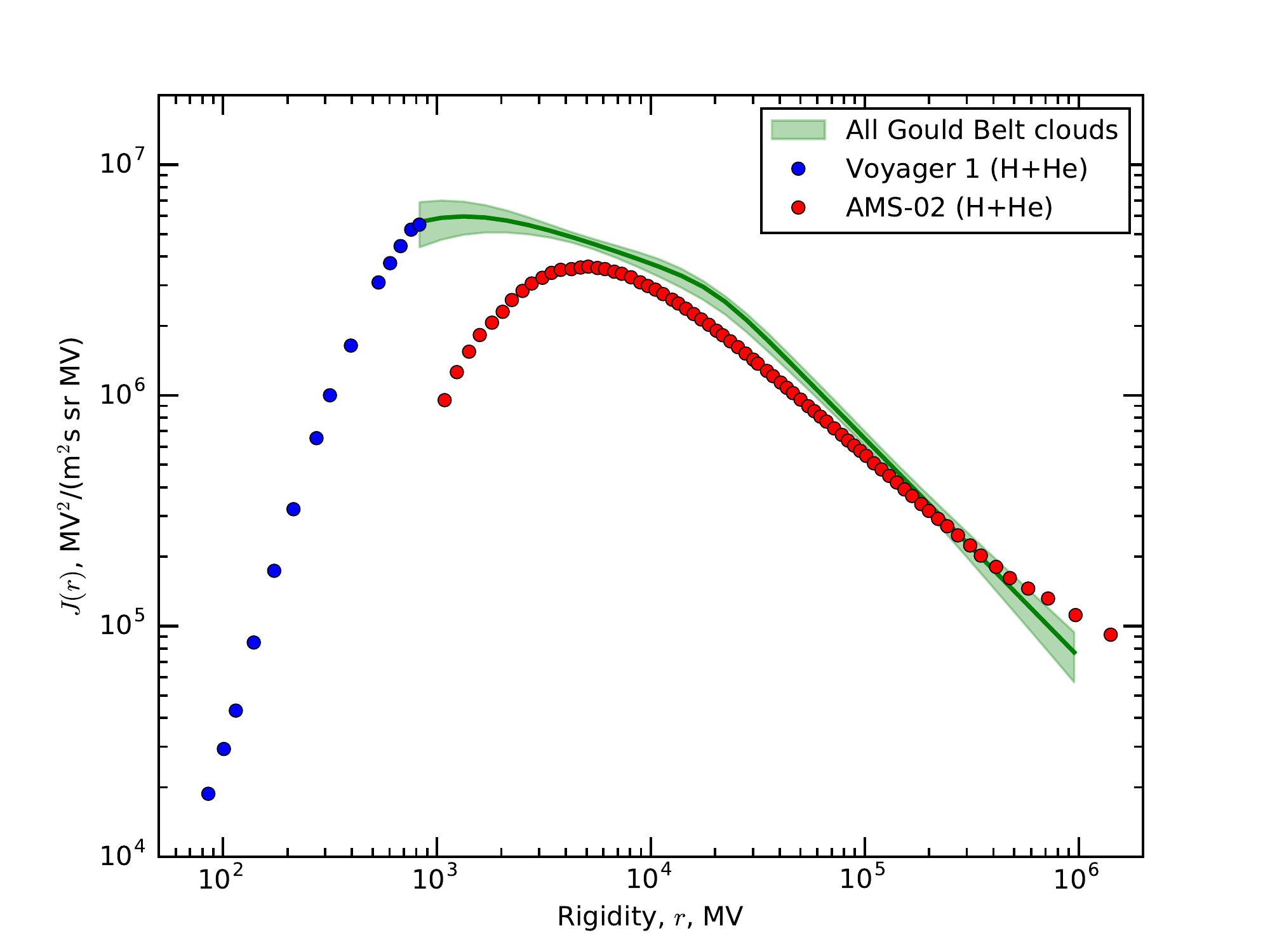}
\includegraphics[width=0.5\linewidth]{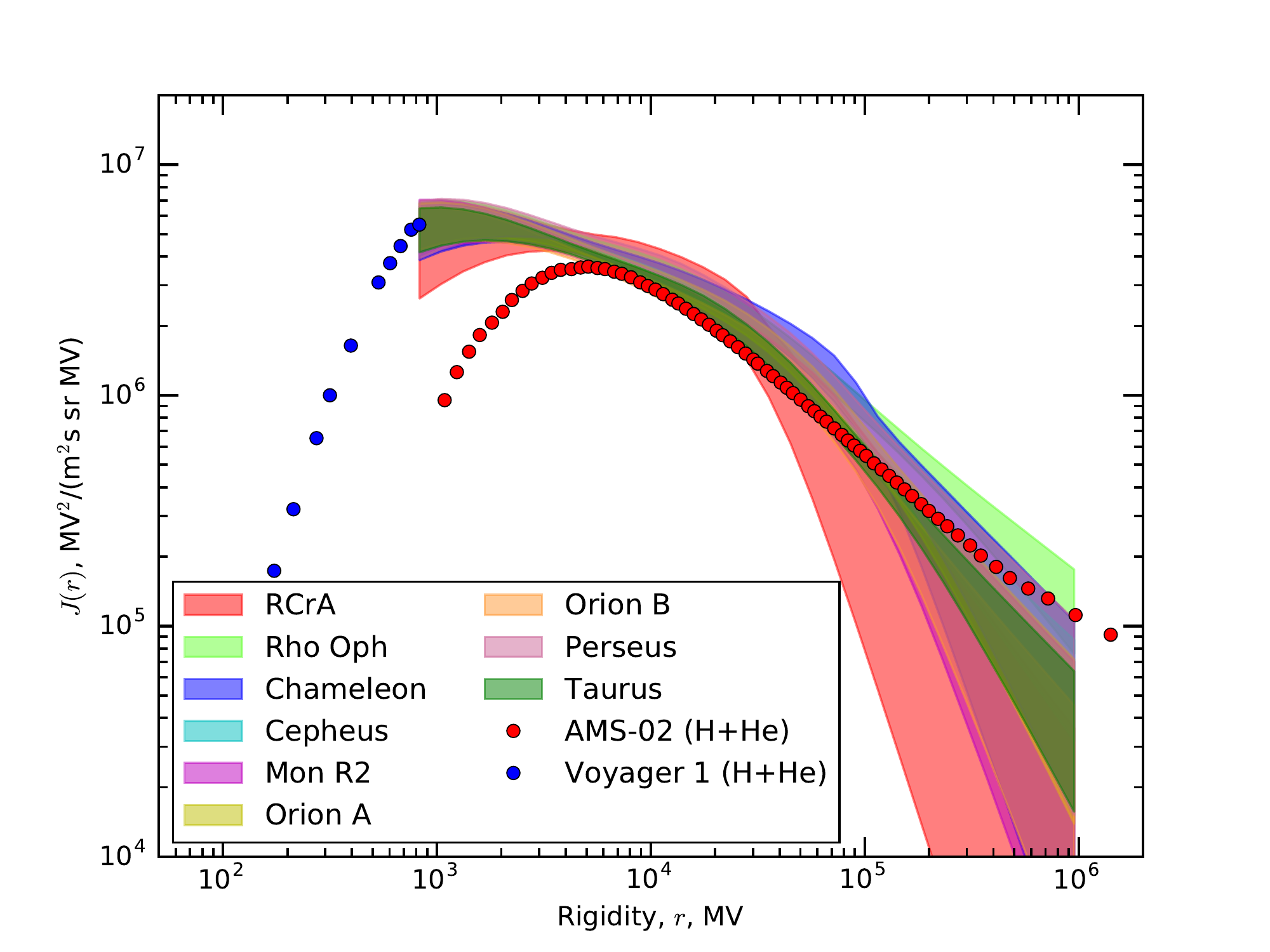}
\caption{Spectrum of cosmic-ray protons and helium measured by Voyager in the interstellar medium below 1~GV \citep{cummings} and by the AMS-02 near Earth above 1~GV \citep{ams-02_proton}. \textit{Left:} the green curve and the shaded band show the broken power-law spectrum in the kpc-sized region around the solar system derived from the Gould Belt observations, normalised on the highest energy Voyager data point. \textit{Right:} spectra of individual Gould Belt clouds normalised to match the highest energy data
point of Voyager.}
\label{fig:cr_spectrum}
\end{figure*}

The \gr\ ``sampling'' of the cosmic-ray spectrum across a region
of 1~kpc in the local Galaxy is also consistent with the measurements near and inside of the solar system. This is illustrated in Fig. \ref{fig:cr_spectrum}, where the \gr\ measurement is compared with combined Voyager and AMS-02 spectral data. In order to account for the contribution of heavier nuclei in the cosmic-ray spectrum, we included in the shown data the contributions from protons and helium ($\sim 10$\% of the flux), added as functions of the rigidity.

The \gr\ measurement is normalised on the Voyager point with
highest energy\footnote{The cosmic-ray spectrum much below 1 GeV is affected by ionisation loss. This explains the gradual hardening of the spectrum at lower energies as measured by Voyager, see e.g.~\citet{webber15,ptuskin15}.}. With this normalisation, the cosmic-ray flux averaged over the 1~kpc region clearly matches the local measurement by AMS-02 and PAMELA at around 200~GV. 

Voyager measurements provide an independent confirmation of the low-energy break in the cosmic-ray spectrum. Figure  \ref{fig:cr_spectrum}  shows that a power-law extrapolation of the AMS-02 or PAMELA spectrum with the slope $2.8 ... 2.85$ toward lower energies would over-predict the Voyager measurement. This has previously
been noted in a number of publications on modelling the combined Voyager and$\text{}$ AMS-02/PAMELA data.  For example, modelling the combined Voyager and PAMELA spectrum by \citet{cummings} considered a range of breaks in the spectrum in the energy range between 1 and 20 GeV, introduced either in the injection spectrum of cosmic rays or in the energy dependence of the cosmic-ray diffusion coefficient. In particular,  the ``diffusive reacceleration'' (DR) model introduces a break  at 
$r_{br,DR}=18\mbox{ GV}$, with a slope change by  $\Delta i=0.54$. This is close to the measurement derived from the Gould Belt data:
\begin{equation}
r_{br}=18.0^{+7}_{-4} \mbox{ GV}, \Delta i=0.59\pm 0.11 
.\end{equation}

The locally measured slope of the spectrum above the break, $i_2$, is somewhat harder than the slope derived from the \gr\ data ($i_2\simeq 2.85$ for PAMELA data \citep{pamela_proton} and $i_2\simeq 2.8$ for the AMS-02 \citep{ams-02_proton} than the $i_2=2.92_{-0.04}^{+0.07}$ for the Gould Belt clouds.  In the Gould Belt clouds, the measurement
statistics of the \gr\ spectrum above hundred GeV is still low, which affects the measurement precision. This is particularly clear in the right panel of Fig. \ref{fig:cr_spectrum}, which shows the range of uncertainties of the high-energy slopes derived from the spectra of individual molecular clouds. 
%%%%%%%%%%%%%%%%%%%%%%%%%%%%%%%%%%%%%%%%%%%%%%%%%
\begin{table*}
\begin{center}
\begin{tabular}{|c|c|c|c|c|c|}
\hline
Name  & $N_0, 10^{44}$~V$^{-1}$ & $i_1$ & $r_{br}$, GV & $i_2$ & s \\
\hline
&&&&&\\[-0.25cm]
R CrA  & $0.24^{+0.04}_{-0.06}$ & $2.33^{+0.08}_{-0.21}$ & $33.72^{+17.33}_{-11.02}$ & $4.82^{+0.11}_{-0.88}$ & 16.06 ( $>$1.03 ) \\[0.1cm]
Rho Oph  & $2.44^{+0.35}_{-0.25}$ & $2.31^{+0.08}_{-0.09}$ & $17.72^{+21.49}_{-4.94}$ & $2.78^{+0.17}_{-0.05}$ & 20.61 ( $>$0.84 )\\[0.1cm]
Perseus   &$1.21^{+0.18}_{-0.14}$ & $2.29^{+0.08}_{-0.11}$ & $20.75^{+32.81}_{-5.77}$ & $2.95^{+0.42}_{-0.07}$ & 9.55 ( $>$0.88 )\\[0.1cm]
Chameleon   & $1.13^{+0.13}_{-0.14}$ & $2.33^{+0.06}_{-0.11}$ & $32.75^{+47.33}_{-10.00}$ & $3.07^{+0.75}_{-0.14}$ & 11.19 ( $>$0.88 )\\[0.1cm]
Cepheus   &$3.97^{+0.43}_{-0.42}$ & $2.36^{+0.06}_{-0.10}$ & $18.06^{+13.10}_{-4.24}$ & $2.92^{+0.18}_{-0.05}$ & 71.02 ( $>$1.02 ) \\[0.1cm]
Taurus   & $5.40^{+0.53}_{-0.54}$ & $2.38^{+0.06}_{-0.09}$ & $21.87^{+19.36}_{-4.33}$ & $3.02^{+0.28}_{-0.06}$ & 56.46 ( $>$1.05 )\\[0.1cm]
Orion A   & $2.54^{+0.32}_{-0.23}$ & $2.35^{+0.07}_{-0.08}$ & $27.03^{+31.30}_{-5.58}$ & $3.05^{+0.38}_{-0.07}$ & 230.94 ( $>$1.00 )\\[0.1cm]
Orion B   & $2.73^{+0.25}_{-0.25}$ & $2.41^{+0.05}_{-0.08}$ & $30.52^{+32.24}_{-6.64}$ & $3.19^{+0.53}_{-0.10}$ & 17.90 ( $>$1.09 ) \\[0.1cm]
Mon R2   &$0.54^{+0.08}_{-0.06}$ & $2.38^{+0.08}_{-0.11}$ & $22.47^{+51.55}_{-6.14}$ & $3.02^{+0.76}_{-0.10}$ & 89.20 ( $>$0.80 )\\[0.1cm]
\hline
&&&&&\\[-0.25cm]
All & $19.41^{+2.11}_{-1.87}$ & $2.33^{+0.06}_{-0.08}$ & $18.35^{+6.48}_{-3.57}$ & $2.92^{+0.07}_{-0.04}$ & 62.52 ( $>$1.50 )\\
\hline
\end{tabular}
\end{center}
\caption{Best-fit parameters for the proton distributions (see Eq.~\ref{eq:fit_model}) in Gould Belt clouds. For the break strength, the best-fit value and 95\% lower limits are given.  Note that the normalisations $N_0$ are scaled to the distance to the cloud $D=1$~kpc, and the cloud density is $n=1$~cm$^{-3}$.}
\label{tab:fit_params}
\end{table*}
%%%%%%%%%%%%%%%%%%%%%%%%%%%%%%%%%%%%%%%%%%%%%%%%%
Below the break, the measured value of the slope, $i_1=2.33^{+0.06}_{-0.08}$, is consistent with the measurement of the average slope of the cosmic-ray spectrum in the inner Galactic disk and in the Large Magellanic Cloud (LMC), as shown in Fig. \ref{fig:best_fit_params}. The value $i\simeq 2.3 ... 2.4$ naturally arises in the simplest model of cosmic-ray propagation in the turbulent magnetic field with a Kolmogorov turbulence spectrum~\citep{kolmogorov41}. If the injection spectrum has a slope $i_0\simeq 2.0 ... 2.1$, as predicted by the diffusive shock acceleration models~\citep[see e.g.][]{fermi49,blandford78,drury83,blandford87,berezhko97}, the steady-state spectrum of cosmic rays propagating through  turbulent magnetic fields has a slope that is softer than the injection spectrum by $\text{one-third}$, that is, $i_1\simeq i_0+1/3\simeq 2.3 ... 2.4$ \citep[see e.g.][ for recent reviews and references therein]{schlickeiser02,aharonian12}.

The similar slopes at low ($2.37 \pm 0.09$) and high ($2.82 \pm 0.05$) energies were also derived for local interstellar gas, see~\citet{dermer13,strong15}. The break energy observed in this case is somewhat lower ($6\pm2$~GeV) than observed in Gould Belt clouds. This might be explained by \fermi/LAT calibration changes and/or by the constrained fit for the energy break that was allowed to vary only in 1-10~GeV range in~\citet{strong15}.

The consistency of the slopes of the average cosmic-ray spectrum in the inner Galactic disk and in the local interstellar medium shows that the propagation mechanisms of cosmic rays are the same across the Galactic disk. Although the strength of the turbulent magnetic field might change at different locations, its turbulent structure is perhaps universal, so that the relation between the injection and the steady-state spectra of cosmic rays is the same everywhere in the disk. This conclusion is opposite to the model assumption of \citet{gaggero}, where the difference between the slopes of the cosmic-ray spectrum in the inner Galaxy and the local interstellar medium was attributed to a gradual change in the energy dependence of the cosmic-ray diffusion coefficient (and hence the change in the turbulence power spectrum of magnetic field)  with increasing distance from the Galactic centre. 

Identical slopes of the local and average Galactic cosmic ray spectra naturally occur in a steady-state regime in which cosmic rays are continuously injected with a nearly constant rate across the interstellar medium, both in the solar neighbourhood and in the inner Galaxy. This approximation is valid when very many sources contribute to the cosmic-ray flux measurement. This is certainly true up to very high energies in the context of  deriving the cosmic-ray spectrum in the inner Galactic disk from the \gr\ flux of the entire disk within the solar circle. However, it is less obvious when the spectral measurements are taken at isolated fixed points, such as the position of the Sun, or when the measurements
are made in individual molecular clouds. 
%%%%%%%%%%%%%%%%%%%%%%%%%%%%%%%%%%%%%%%%%%%%%%%%%
%%%%%%%%%%%%%%%%%%%%%%%%%%%%%%%%%%%%%%%%%%%%%%%%%
When the sources injecting cosmic rays are supernovae, the assumption about a continuous and constant rate injection does not necessarily hold for single fixed-point measurements.  Supernovae inject cosmic rays at discrete moments of time, with approximately one injection episode per Myr per (100 pc$)^2$ patch of the Galactic disk.  Cosmic rays spread in the interstellar medium in an anisotropic way, with a much slower diffusion across the magnetic field lines than along the magnetic field \citep{casse02,demarco07,giacinti12}. For a local Galactic magnetic field of the strength $B\simeq 4\ \mu$G with moderate turbulence level $\eta=B_{turb}^2/(B_{turb}^2+B_{ord}^2)\sim 0.1$ ($B_{turb}, B_{ord}$ being the turbulent and ordered components of the field), the diffusion coefficient perpendicular to the field is two orders of magnitude smaller than the coefficient parallel to the field, $D_\bot\sim 10^{-2}D_\|$. The cosmic rays  escape into and from the kiloparsec-scale  cosmic-ray halo  around the disk as a result of the non-zero vertical component of the ordered magnetic field  $B_{ord,z}$  \citep{farrar,jansson12}. By the time that cosmic rays are spread over the entire thickness of the halo, $H$, in the vertical direction, they still occupy a thin tube of the length  $L=H B_{ord}/B_{z,ord}$  and thickness  $l=(D_\bot/D_{\|})L\sim 10^{-2}L$ along the ordered field line. The thickness of this tube is small, only about $l\sim 100$~pc, even when the tube length is  in the $L\sim 10$~kpc range. 

This anisotropic diffusion picture is dramatically different from the  isotropic diffusion considered as a baseline for the leaky-box model and from numerical codes like GALPROP \citep{galprop}.  In the isotropic diffusion picture ($D_\bot\sim D_\|$), each supernova spreads cosmic rays into the kiloparsec-scale volume all around the source. These kiloparsec-scale volumes around different supernovae overlap so that thousands of supernovae could simultaneously contribute to the GeV energy band cosmic-ray  flux at any given point. In contrast, very slow diffusion in the direction perpendicular to the ordered magnetic field strongly reduces the volume occupied by cosmic rays that are spread by individual supernovae in the anisotropic diffusion picture. Only some tens of supernovae contribute to the GeV band flux at any given point. 

The escape time of cosmic rays decreases with the energy increase. This leads to a further decrease in the number of supernovae contributing to the flux at higher energies.  If the escape time becomes comparable to the interval between subsequent injection episodes into the narrow tubes, the fluctuations of the cosmic ray spectrum become large and the steady-state approximation is not appropriate to describe the spectrum in each tube.

The last injection in the local interstellar medium has apparently occurred some 2~Myr ago \citep{local_source,savchenko15} and has left a trace in the primary cosmic-ray particle spectrum in the TeV energy range as well as in the spectra of positrons and antiprotons in the 100~GeV range. The location of the solar system is not special. TeV-range cosmic-ray spectra at different locations across the Gould Belt are probably also influenced by the last supernovae local to these locations. The timing of the last supernovae at different places is not synchronised, and the influence of the last supernova on the cosmic-ray spectra in different clouds is also expected to be different. This naturally leads to the variations in TeV cosmic-ray fluxes across the local interstellar medium, or, in other words, to the variations in the slope $i_2$ of the cosmic-ray spectrum in the range of 10~GeV to TeV. 

The statistics of Fermi/LAT data is only marginally sufficient for measuring this effect. We do find significant differences between the high-energy slopes of the cloud spectra (with the identical slope hypothesis inconsistent with the data at the $2.7\sigma$ level).  The most significant deviation of the high-energy slope from the cloud-average slope is found  in the RCrA cloud. However, the RCrA has the lowest \gr\ flux of all the analysed clouds. It is also found in the direction of the Fermi bubbles above/below the Galactic centre where the modelling of diffuse Galactic background suffers from large ambiguities. This cloud was also found to have a peculiar spectrum in the analysis of \citet{yang14}, possibly contaminated by the hard spectrum of Fermi bubbles, which is due to difficulties in modelling the diffuse background. When we removed this cloud from the sample, we found that the hypothesis of an identical high-energy slope in all clouds became more consistent with the data (inconsistent at the $2.3\sigma$ level). However, the measurement of the higher-energy slope averaged over a region of 1 kpc is inconsistent with the local measurement of the spectral slope in the 20-200 GV range by AMS-02 and PAMELA. This indicates either that the solar system is located in a special place (e.g. the environment is influenced by the Local Hot Bubble) or that the slope of the cosmic-ray spectrum above the break is generically variable across the entire 1 kpc region. 

The discreteness of injection events of cosmic rays in time/space is less important at lower energies where 
 the number of supernovae contributing to the cosmic-ray flux at a given point (section of a tube along the ordered magnetic field direction)  is much larger than one because the residence time of cosmic rays is much longer. It might be expected that in this case the steady-state approximation of an injection continuous in time/space provides an appropriate model for describing the cosmic-ray content of the local interstellar medium. This seems to be the case in the energy range below 20~GeV,  in which the spectral slopes of different clouds are consistent with each other.  A softening of the spectrum above $r_{br}\simeq 18$~GV is then attributed to the gradual transition from the steady-state continuous injection to the regime of discrete source injection.
 
An alternative  explanation for the break could be a variability of the local star formation rate on the 10 Myr  escape timescale of 10~GV cosmic rays. The solar system is situated inside the Local Hot Bubble, a cavity carved out by subsequent supernovae over the past 10-20 Myr  \citep{localbubble}.  On a somewhat longer timescale, the star formation in our Galactic neighbourhood was shaped by the activity of regions that are now forming the Gould Belt itself \citep{gouldbelt}. The cosmic-ray flux at different energies is proportional to the average star formation/supernova rate in the local Galaxy averaged over the cosmic-ray escape time. The different escape times at different cosmic ray energies lead to the variations in flux normalisation.  In this scenario there is no physical explanation for the consistency between the slope of the local cosmic-ray spectrum below the break and that of the average Galactic cosmic-ray spectrum. This consistency needs to be considered as a coincidence.   

Still another possible explanation for the break might be the change in the mechanism of the cosmic-ray spread through the interstellar medium, for instance, the transition between the convective and diffusive propagation regimes \citep{blasi12,nath12}. 
However,  in this case, the break would be also identifiable in the energy dependence of the primary-to-secondary cosmic-ray ratio (like B/C). No particular feature is detected at this energy \citep{pamela_bc,ams-02_bc}. Thus, this explanation is not consistent with the data. 
 
%%%%%%%%%%%%%%%%%%%%%%
\section{Conclusions}
%%%%%%%%%%%%%%%%%%%%%%

The updated analysis of the \gr\ signal from the Gould Belt molecular clouds confirms our previous finding \citep{lowenergybreak} that a low-energy break exists in the spectrum of cosmic rays in the local Galaxy (averaged over a region of 1 kpc). This break was now also directly established by combining the local measurements of the cosmic-ray spectrum in the interstellar medium by Voyager. 

The slope of the cosmic-ray spectrum below the break $i_1=2.33_{-0.08}^{+0.06}$ is close to the average slope of the cosmic ray spectrum in the inner disk of the Milky Way (see Fig. \ref{fig:best_fit_params}). This value of the slope is consistent with the model of cosmic-ray injection and propagation in the Galaxy in which cosmic rays are injected with a standard slope $2.0 ... 2.1$ by the shock-acceleration process and escape through the turbulent magnetic field with the Kolmogorov turbulence spectrum. 

The slope of the spectrum above the break is found to be variable across the 1~kpc scale region. The slope variability might be due to the discreteness of cosmic-ray injection events in space and time. This possibility is valid when cosmic-ray diffusion proceeds in a strongly anisotropic way. This scenario also provides consistent explanations for the slope variation of the locally measured cosmic-ray spectrum at higher energies (hundreds of GeV) where contributions from individual cosmic-ray sources (like
a single recent supernova) become increasingly more important \citep{local_source,savchenko15}.

\noindent \textit{Acknowledgements.} 
The work of DM was supported by the Carl-Zeiss Stiftung through the grant ``Hochsensitive Nachweistechnik zur Erforschung des unsichtbaren Universums'' to the Kepler Center f{\"u}r Astro- und Teilchenphysik at the University of T{\"u}bingen.

%\bibliography{low_energy_break}

\end{document}